\documentclass[twocolumn,amsmath,amssymb,prl,a4paper]{revtex4}
\usepackage{graphicx}

\newcommand{\tr}{\mathop\mathrm{tr}}

\newcommand{\ket}[1]{\left|#1\right>}

\newcommand{\expc}[1]{\left<#1\right>}

\newcommand{\hB}{{\hat B}}

\date{\today}
\begin{document}
\title{Wave packet Rabi oscillations from phase-space flow in mesoscopic cavity QED}
\author{Omri Gat}
\affiliation{Racah Institute  of  Physics, Hebrew University
of Jerusalem, Jerusalem 91904, Israel}
\begin{abstract}
The dynamics of the Jaynes-Cummings Hamiltonian is studied in the semiclassical limit by reducing it to quantum normal form and deriving the phase space dynamics of classically smooth observables. The resulting systematic semiclassical approximation is used to analyze the Rabi oscillations of a photon wave packet revealing subtle consequences of the uncertainty principle.

\end{abstract}
\maketitle
\paragraph{Introduction}
The Jaynes-Cummings (JC) Hamiltonian of a two-state system coupled to a harmonic oscillator is a basic model of quantum optics and the theory of atom-light interactions. This simple Hamiltonian describes quite accurately the quantum electrodynamics of strongly coupled Rydberg atoms in ultra-high Q microwave cavities \cite{haroche-rmp}.
Because of the universal nature of the JC Hamiltonian, it models many other quantum system with interactions between discrete and continuous degrees of freedom, like hyperfine levels in trapped ions \cite{trapped} and spin-orbit interactions in semiconductors \cite{rashba}. It is also a building block for more complicated systems like micromasers \cite{micromasers} and electromagnetically induced transparency cavities \cite{eit}.

In the rotating wave approximation (RWA), where non-resonant terms in the Hamiltonian are neglected, the JC model becomes integrable, with a conserved quantity in addition to energy --- the polariton number that gives the total number of excitations, field and atom. Polariton number conservation makes it easy to diagonalize the Hamiltonian, since the polariton eigenspaces are only two-dimensional, and makes the dynamics of states with a small photon number simple. Nevertheless, the dynamics of states that have many significant energy-state components, including the important example of phase-space localized wave packets, can show considerable complexity \cite{larson}, including squeezing \cite{squeezed}, wave packet Rabi oscillation, collapse-revival \cite{eberly}, and entanglement \cite{gea}.

Recently the interest in wave packet dynamics in the JC system has increased, as it has become a subject of direct experimental investigation in cavities excited to  mesoscopic coherent states with a moderate number of photons \cite{haroche-mesoscopic}. The experimental systems are of course also subject to dissipation and decoherence, but the coherence lifetimes that have been achieved ensure that the actual dynamics can be approximated by that of an ideal system during the initial time interval of wave packet evolution, where the complex dynamical phenomena occur.

Complex wave packet dynamics can be exhibited by the integrable JC system because its energy spectrum is a nonlinear function of the number of polaritons. Many of the complex dynamical phenomena occur also in the simpler case of nonlinear oscillators \cite{gat-nlo,berry}, where they can be analyzed using semiclassical nonlinear dynamics. Here we show that an analogous semiclassical analysis is applicable also to the JC dynamics. 

As always, the semiclassical approximation is valid when the typical action scales are significantly larger than Planck's constant. In the JC system this condition holds either when the intrinsic action scale $B_R$ (defined below), that is a measure of the detuning, is much larger than $\hbar$, or when the initial state is sufficiently excited energetically. Then the field degree of freedom becomes semiclassical but,
unlike nonlinear oscillators, the JC system also has the atomic polarization internal degree of freedom that remains fully quantum mechanical. This contrast is reflected by the much faster dynamics of the polarization versus the field: The ratio of the natural frequencies of the polarization and the field dynamics diverges in the semiclassical limit. Our semiclassical analysis, that retains an exact quantum-mechanical treatment of the internal degree of freedom, is in this way different from that of cavity dynamics of an ensemble of atoms \cite{haken}.

Wave packet propagation studies \cite{eberly,gea} have provided a semiclassical analysis of the JC system maintaining a quantum internal degree of freedom, but their utilization of the classical dynamics is limited because Scr\"odinger picture wave packet propagation does not have a direct classical limit. Here, in contrast,
the dynamics is obtained by evolving classically smooth observables in the Heisenberg picture. The dynamics of the observables is cast in terms of their semiclassical phase-space \emph{quasi-flow} \cite{gat-nlo,osborn-robert}, defined as their time-dependent Weyl representation. As a consequence of the presence of internal degrees of freedom the quasi-flow of the JC system is  \emph{matrix-valued}, giving the components of an operator in polarization subspace at each phase-space point.

\begin{figure}[htb]

\includegraphics[width=7.cm]{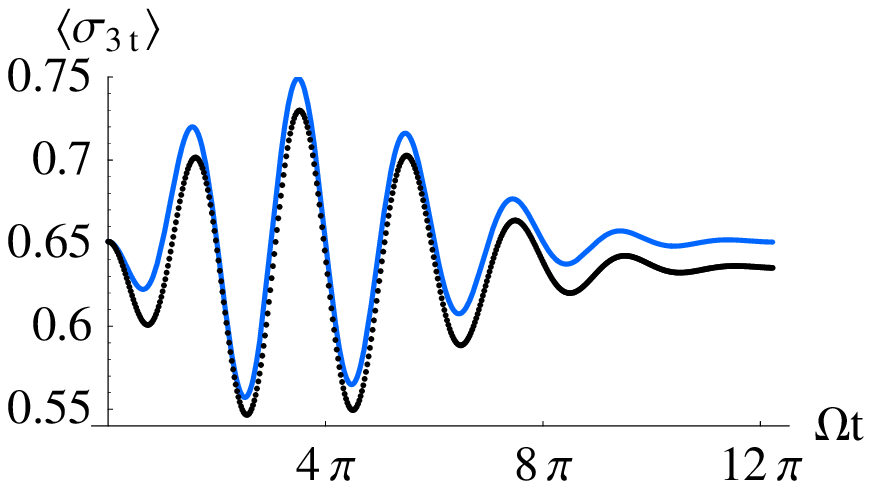}\break\includegraphics[width=7.cm]{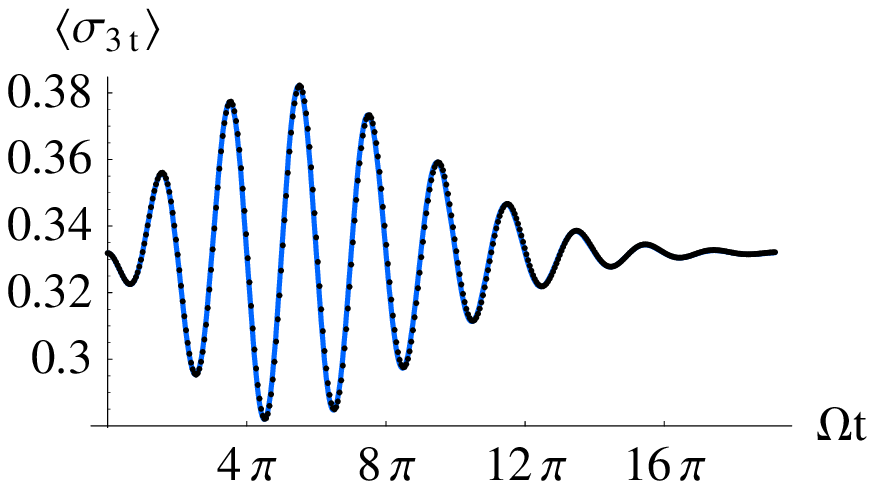}
\caption{\label{fig:dark} (Color online) The mean polarization of an initially coherent photon state, and atomic polarization optimally localized on the `+' invariant subspace, with $B_R=6.25\hbar$. The continuous blue line is derived from the semiclassical approximation Eq.\ (\ref{eq:sigma3t}) and the black dots from a direct numerical solution. Good agreement is observed already in the top panel where the mean polariton number is 8 while the bottom panel with 50 polaritons shows agreement to less than 3\%. The relative error term in the semiclassical approximation is $O(\hbar^{1/2})$.
}
\end{figure}

\begin{figure}[htb]
\includegraphics[width=7cm]{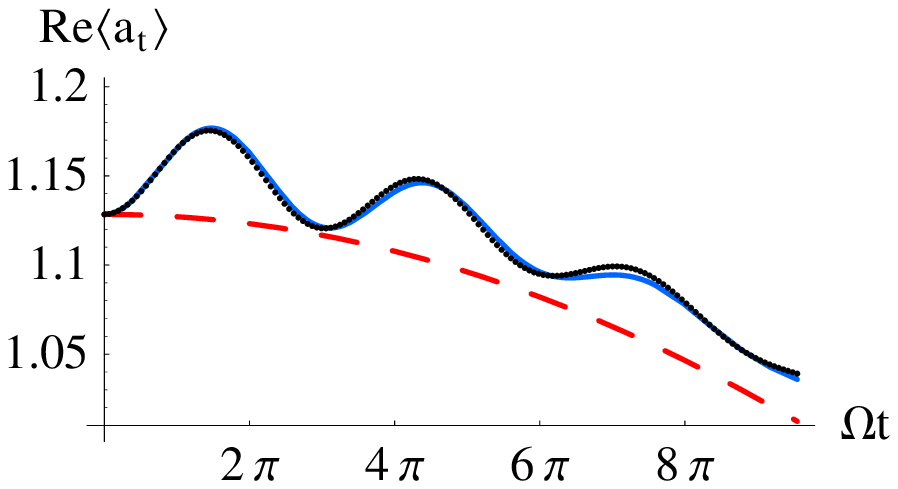}\break\includegraphics[width=7.cm]{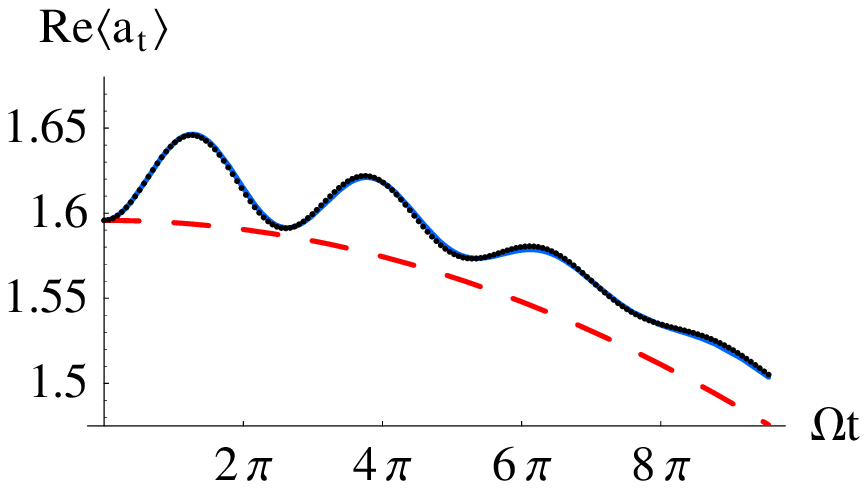}
\caption{\label{fig:field} (Color online) The expectation of the real quadrature of the field amplitude of an initially coherent photon and excited atomic state with $B_R=6.25\hbar$. The continuous blue line and the black dots have analogous meanings to those of Fig.\ \ref{fig:dark}, while the dashed red line shows only the adiabatic terms in the semiclassical field dynamics Eq.\ (\ref{eq:at}). Good agreement is observed for the lower polariton numbers, 4 (top) and 8 (bottom) since the relative semiclassical error  is $O(\hbar)$.
}
\end{figure}

The focus of this work is the Rabi oscillations of wave packets. The semiclassical quasi-flow leads to the explicit calculation of the Rabi dynamics beyond the leading-order classical substitution, demonstrating subtle quantum effects in the interaction between the slow and fast degrees of freedom: The entanglement of eigenstates results in the polarization oscillations shown in Eq.\ (\ref{eq:sigma3t}) and Fig.\ \ref{fig:dark}, while the quantum field uncertainty causes the \emph{field} Rabi oscillations shown in Eq.\ (\ref{eq:at}) and Fig.\ \ref{fig:field}.

\paragraph{The polariton quasi-flow}
The JC model describes a single electromagnetic mode of frequency $\omega$, modeled by a harmonic oscillator with a lowering operator $a$, interacting with a two-state system of energy gap $\hbar\nu$ with a lowering operator $\sigma$. Under the RWA, where the non-resonant terms $a\sigma$ and $a^\dagger\sigma^\dagger$ are neglected, the JC Hamiltonian is
\begin{equation}\label{eq:jcm}
H=\hbar\omega\frac{a^\dagger a+a a^\dagger}{2}+\hbar\nu\frac{\sigma^\dagger \sigma-\sigma \sigma^\dagger}{2}+\hbar g(a^\dagger\sigma+a\sigma^\dagger)\ .
\end{equation}
When the two-level system models an atom, $g$ is an effective dipole interaction. The commuting $a$ and $\sigma$ operators obey Bose and Fermi algebra rules, respectively, i.e., $[a,a^\dagger]=1$, $\sigma^2=0$, $[\sigma,\sigma^\dagger]_+=1$. 

The RWA JC Hamiltonian is integrable because it commutes with the \emph{polariton} number operator $n_{\text{pol}}=a^\dagger a-\sigma^\dagger \sigma$. In other words, the system excitations are polaritons     \cite{rebic} with an internal two-state polarization degree of freedom. For the purpose of studying dynamics we therefore define the (single-mode) bosonic polariton annihilation operator 
$b=\sqrt{\hbar}(\sigma^\dagger\sigma+\sigma\sigma^\dagger\sqrt{\frac{a^\dagger a}{aa^\dagger}})a$, and the fermionic polarization lowering operator $\tau=\sigma a^\dagger\frac{1}{\sqrt{aa^\dagger}}$ that commutes with it. That is, $b^\dagger$ and $b$ create and annihilate (respectively) polariton quanta, and $\tau^\dagger$ and $\tau$ raise and lower (respectively) the polarization. We define also standard $SU(2)$ generators for the internal degree of freedom of the polaritons, $\tau_1=\tau+\tau^\dagger$, $\tau_2=i(\tau-\tau^\dagger)$, $\tau_3=\tau^\dagger\tau-\tau\tau^\dagger$.

In terms of $b$ and $\tau$ the Hamiltonian takes the form
$
H=\omega\hB +\delta \tau_3+\lambda\sqrt\hB\tau_1
$,
where $\delta=\frac\hbar2(\nu-\omega)$, $\lambda=\sqrt{\hbar}g$~, $\hB=bb^\dagger=\hbar(n_{\text{pol}}+1)$, and $\sqrt{\hB}$ is the unique positive operator whose square is $\hB$. The advantage of the polariton representation is that the Hamiltonian depends on the polariton field only through the combination $\hB$; that is, the Hamiltonian has been reduced to its quantum normal form \cite{lj}\footnote{The normal form Hamiltonian  is equal to the original Hamiltonian (\ref{eq:jcm}) except for its action on the ground state. Rather than add an encumbering correction term to the Hamiltonian it will be assumed that there is a negligible probability amplitude to be in the ground state, a natural assumption in the semiclassical analysis.}. 
At this stage $\hB$ can be handled as if it were a c-number, letting the Hamiltonian be further reduced by invariant polarization space decomposition into
$H=P_+(\hB)\varepsilon_+(\hB)+\hat P_-(\hB)\hat\varepsilon_-(\hB),$
introducing the definitions $\varepsilon_\pm(\hB)=\omega\hat B\pm\varepsilon(\hB)$, $\varepsilon(\hB)=\lambda\sqrt{\hat B+B_R}$, where $B_R=(\delta/\lambda)^2$ is the natural action scale of the system. $\varepsilon_\pm(\hB)$ are scalars in the internal degree of freedom space, i.e.\ they commute with $\tau$ and $\tau^\dagger$.
$P_\pm(\hB)=\frac12(1\pm(s(\hB)\tau_1+c(\hB)\tau_3))$ are orthogonal projections such that $P_+ + P_-=1$, $ P_+ P_-=P_- P_+=0$, with $c(\hB)=\sqrt{B_R/(\hB+B_R)}$, and $s=\sqrt{1-c^2}$.
The $P_\pm(\hB)$ are projections on invariant \emph{dressed} polarization subspaces that, unlike the bare polarization projections $\frac12(1\pm\tau_3)$, do not commute with the field $b$.

The reduction of the Hamiltonian to its normal form simplifies the solution of the Heisenberg equations of motion. The dynamics of the polarization operator $\tau_3$ is given by 
\begin{align}\nonumber
\tau_{3,t}&=\bigl(s(\hB)\tau_1+c(\hB)\theta\tau_3\bigr)c(\hB)- \bigl(s(\hB)c(\hB)\tau_1-\tau_3\bigr)\\&\times\cos\bigl(\Omega(\hB) t\bigr) + s(\hB)\sin\bigl(\Omega(\hB) t\bigr)\ ,
\label{eq:taut}\end{align}
where $\Omega=\frac2\hbar\varepsilon$ is the  Rabi frequency.
The dynamics of the polariton field $b$ is derived using the identity $b f(\hB)=f(\hB+\hbar)b$ that holds for an arbitrary function $f$,
\begin{align}\label{eq:bt}
b_t&=\bigl(P_+(\hB)+P_-(\hB)e^{-i\Omega(\hB)t}\bigr)P_+(\hB+\hbar)e^{-i\tilde\omega_\hbar(\hB)t}
\\\nonumber&+\bigl(P_+(\hB)e^{i\Omega(\hB)t}+P_-(\hB)\bigr)P_-(\hB+\hbar)e^{i\tilde\omega_\hbar(\hB)t})e^{-i\omega t}b\ ,
\end{align}
defining the (field-dependent) phase frequency
$\tilde\omega_\hbar=\frac1\hbar(\varepsilon(\cdot+\hbar)-\varepsilon(\cdot))$. $\tilde\omega_\hbar$ is semiclassically slow and has a well-defined value $\tilde\omega=\varepsilon'$ in the classical limit $\hbar\to0$.

The dynamical analysis presented so far is exact, but the conversion of Eqs.\ (\ref{eq:taut}) and (\ref{eq:bt}) to physically measurable quantities requires taking matrix elements with respect to the initial state, and for this purpose the semiclassical approximation will be assumed, letting $\hbar\to0$ keeping other system parameters fixed. The analysis will rely on the classical smoothness of $\tau_{3,t}$ and $b_t$, and this requires the time $t$ to be of $o(1)$ in the semiclassical limit. As shown below, the time of validity encompasses many Rabi oscillations and their collapse but not their revival.

The semiclassical approximation will be made in the Wigner-Weyl representation of the JC system that is a special case of the standard Wigner-Weyl representation of many-component systems \cite{lj-f}, in itself a straightforward generalization of the scalar Wigner-Weyl representation. Because the polarization degree of freedom remains quantum-mechanical, the symbol range consists of the operators in the two-dimensional polarization Hilbert space. In the same manner, the Wigner ``function'' $W$ is also operator-valued and
the expectation value of an observable is
\begin{equation}\label{eq:phyex}
\expc{\hat L}=\int\frac{d^2\alpha}{2\pi}\tr L(\alpha,\bar \alpha)W(\alpha,\bar \alpha)\ ,
\end{equation}
where $L$ is the Weyl representation of the operator $\hat L$, $\alpha$ is a complex phase-space coordinate, and $\tr$ stands for trace on the two states of the internal degree of freedom.

The dynamical expectation values presented below were calculated using Eq.\ (\ref{eq:phyex}) after deriving the quasi-flow, i.e. the time dependent Weyl representation, of $\sigma_3$ ($=\tau_3$) and of $a$ from Eqs.\ (\ref{eq:taut}) and (\ref{eq:bt}). Because these operators are classically smooth, the leading term quasi-flow is obtained simply by substituting the classical field amplitude $\alpha$ and its complex conjugate (respectively) for the operators $b$ and $b^\dagger$ (note that the difference between the $a$ and $b$ operators is of higher order in $\hbar$). The higher order terms were obtained using the standard rules of the Moyal calculus \cite{lj-f}. The straightforward but technical details of this calculation are omitted.

\paragraph{Semiclassical Rabi oscillations}
We proceed to a concrete demonstration of the results by the Rabi oscillations of phase-space localized pure initial states. Specifically, the initial states are taken to be separable in the \emph{bare} field-atom representation with a coherent field state. This means that the initial  Wigner ``function'' is of the form $2w\exp(-2|\alpha-\alpha_0|^2)$, where $w$ is a projection operator in the atomic two-dimensional Hilbert space. The Wigner function is combined with the quasi-flows to calculate physical expectation values, by carrying out the Gaussian phase-space integral and the two-dimensional trace operation in Eq.\ (\ref{eq:phyex}).

Consider first the case where the atom is prepared in the excited state, $w=\frac12(1+\sigma_3)$, and the observable to be measured is the mean polarization $\expc{\sigma_3}$ that gives the difference between the probabilities that the atom is in the excited state and in the ground state. In a classical field of intensity equal to the mean intensity of the quantum field, the two-level system would undergo periodic Rabi oscillations of frequency $\Omega(A_0)$, where $A_0=\hbar(|\alpha_0|^2+\frac12)$ is the expectation value of the action in the coherent initial state $\ket{\alpha_0}$. However, because of the field uncertainty, the Rabi oscillations have a finite lifetime \cite{eberly} after which they collapse.
The semiclassical calculation agrees with this result:
\begin{equation}\label{eq:collapse}
\left<\sigma_{3t}\right>=c(A_0)^2+s(A_0)^2\cos(\Omega(A_0) t)e^{-\frac12\hbar A_0(\Omega'(A_0)t)^2}
\end{equation}
exhibits Rabi oscillations of frequency $\Omega(A_0)$, super-exponentially damped with the rate $\sqrt{\hbar A_0}\Omega'(A_0)$. The collapse time is therefore asymptotic to $\hbar^{-1/2}$; in particular it diverges in the classical limit as expected. It should be mentioned that Eq.\ (\ref{eq:collapse}) is obtained from Eq.\ (\ref{eq:taut}) using only the leading classical substitution term in the quasi-flow; therefore, the same collapse dynamics would be observed with a random \emph{classical} electromagnetic field with a Gaussian distribution.

A more subtle consequence of the quantum uncertainty in the electromagnetic field is the impossibility of concentrating a state that is separable in the bare atom-field decomposition on one of the $\pm$ dressed polarization subspaces, because the energy eigenstates are entangled. As a result, the dynamics of any separable initial state exhibits Rabi oscillations between the two internal polariton states. The  separable state will be maximally localized on the $+$ subspace if we choose $w=P_+(A_0)$. The mean polarization obtained with this initial state is
\begin{align}\label{eq:sigma3t}
\left<\sigma_{3t}\right>&=c(A_0)-\hbar\frac{\Omega'(A_0)t}{\Omega(A_0)}s(A_0)\bigl(A_0\theta'(A_0)\\\nonumber&+\tfrac12s(A_0)\bigr)\sin\bigl(\Omega(A_0) t\bigr)\exp\bigl(-\tfrac12\hbar A_0(\Omega'(A_0) t)^2\bigr).
\end{align}
Beside the constant term, that is the only one to survive in the classical limit, the polarization displays Rabi oscillations with an amplitude proportional to $\hbar/A_0$, that is {growing in time}. The amplitude does not grow without bound, however, because it is also subject to collapse. Since the collapse time grows as $\hbar^{-1/2}$ in the semiclassical limit, the peak amplitude is of $O(\hbar^{1/2})$. A comparison of the semiclassical calculation with results obtained by direct numerical solution of the Schr\"odinger equation for a particular parameter choice is shown in Fig.\ \ref{fig:dark}, showing good agreement for a moderate number of photons.

A different manifestation of the quantization of the electromagnetic field is the oscillations of the field amplitude $\expc{a}$ with the fast Rabi frequency. One cause for these oscillations is the periodic energy exchange between the field and atom. More surprising, though, are the Rabi oscillations in the \emph{polariton} field, caused by the non-trivial commutator $[b,b^\dagger]$ that generates the   off-diagonal terms in Eq.\ (\ref{eq:bt}).  

As an example consider again the case where the atom is initially in the excited state, $w=\frac12(1+\tau_3)$. In the frame rotating with the optical frequency $\omega$, the expectation value of the field amplitude is
\begin{align}\label{eq:at}
&\left<a_t\right>e^{i\omega t}=\Bigl(\cos\bigl(\tilde\omega(A_0t)\bigr)-ic(A_0)\sin\bigl(\tilde\omega(A_0)t\bigr)\Bigr)\\&\quad-\frac{\hbar}{4A_0}s(A_0)^2\Bigl(1-e^{-i\Omega(A_0) t}\exp\bigl(-\tfrac12\hbar A_0(\Omega' t)^2\bigr)\Bigr)\alpha_0.\nonumber
\end{align}
The terms in the second line express the Rabi oscillations of amplitude proportional to $\hbar/A_0$. The terms on the first line, on the other hand, result from weighted average of the two diagonal elements in Eq.\ (\ref{eq:bt}) that describe adiabatic phase oscillations in the $\pm$ dressed polarization invariant subspaces frequency $\pm\tilde\omega$ (respectively). In the classical substitution approximation these phase oscillations lead to the splitting of the diagonal components of the Wigner function \cite{eberly,gea,haroche-mesoscopic}.

The time evolution of the the field amplitude as obtained from Eq.\ (\ref{eq:at}) is shown in Fig.\ \ref{fig:field} for a particular parameter choice, and compared with the values obtained from the numerical solution. The ``baseline'' of the Rabi oscillations is provided by the adiabatic terms, whose contribution is shown in the figure with a dashed line. The adiabatic motion is so slow, that in the time interval before the collapse of the Rabi oscillations, only a fraction of a phase oscillation is achieved. However, because of the much larger amplitude of the adiabatic motion, its effect is comparable to that of the fast Rabi oscillations, illustrating the singular nature of the semiclassical limit, where phenomena occurring on widely separated time scales have to be considered simultaneously.

\paragraph{Conclusions}
The dynamics of quantum systems, including integrable ones, becomes computationally complex for phase-space localized wave packets, or other initial states having many nonzero energy components. However, this difficulty can be bypassed when expectation values of classically smooth observables are sought, and the utility of this approach for multi-component systems was demonstrated with the semiclassical analysis of the Rabi oscillations in the Jaynes-Cummings system, a phenomenon that is a direct manifestation of its quantum multi-component nature.

Computability is not a practical concern for  the JC system, but the semiclassical analysis provides a thorough and systematic theory of the dynamics that is lacking in pure numerical analysis. The results presented in Eqs.\ (\ref{eq:collapse})--(\ref{eq:at}) illustrate the fact that the Rabi oscillations of a phase-space localized wave packet are approximated in the leading term by oscillations in a {classical} field with parameters determined by the center of the wave packet. This result is the starting point for the analysis of quantum electromagnetic effects as an expansion in powers of $\hbar$. The concept was demonstrated using photon coherent states, but clearly holds for any phase-space localized initial state. Furthermore, the idea is not limited to the analysis of Rabi oscillations: It is applicable to most other quantities of interest in the JC system such as squeezing and entanglement, as well as to other multi-component quantum system including some cited in the introduction.

The approximation of quantum electrodynamic phenomena by classical ones is, like all semiclassical dynamical approximations, nonuniform in time. In our case, the time scale of validity is set by the Heisenberg time $t_H=(\hbar\Omega')^{-1}$, where the discreteness of energy levels becomes manifest with the revival of oscillations \cite{eberly}, that is not perturbative in the quasi-flow.  Neither is the semiclassical quasi-flow well-suited to study quantum interference phenomena. These phenomena are of course amenable to semiclassical analysis, but they require the use of more elaborate tools and considerably heavier analysis, in contrast with the purpose of this work.

\end{document}